\documentclass[aps,twocolumn,pre]{revtex4-1}
\usepackage{graphicx,amsmath, amsfonts,bm, color}

\usepackage{xcolor}
\usepackage[hypertexnames=false]{hyperref}
\hypersetup{
	colorlinks,
	linkcolor={blue!50!black},
	citecolor={blue!50!black},
	urlcolor={blue!80!black}
}

\usepackage{bm}
\usepackage{braket}

\usepackage{amsfonts,amssymb}
\usepackage{epsfig, amsmath}
\usepackage{bm}
\usepackage{soul}
\usepackage{amsmath}
\usepackage{amssymb}
\usepackage{bm}
\usepackage{verbatim}
\usepackage{graphicx}
\usepackage{subfigure}

\usepackage{psfrag}
\usepackage{epsfig}
\usepackage{mathrsfs}

\usepackage{soul}
\DeclareMathAlphabet{\mathitbf}{OML}{cmm}{b}{it}
\newcommand{\B}[1]{{\mathitbf{#1}}}
\newcommand{\calBold}[1]{\mbox{\boldmath${\cal #1}$}}
\newcommand{\mathBold}[1]{\mbox{\boldmath$#1$}}
\newcommand{\C}[1]{{\mathcal{#1}}}
\newcommand{\tr}{\hbox{tr}}
\renewcommand{\=}{\!=\!}
\newcommand{\tripleCdot}{\stackrel{\mbox{\bf\scriptsize .}}{:}}
\newcommand{\dbar}{{\,\mathchar'26\mkern-12mu d}}

\newcommand{\pa}{\partial}
\renewcommand{\=}{\!=\!}

\begin{document}
\title{Local thermal energy as a structural indicator in glasses}

\author{Jacques Zylberg$^{1}$, Edan Lerner$^{2}$, Yohai Bar-Sinai$^{1,3}$ and Eran Bouchbinder$^{1}$}
\affiliation{$^1$Chemical Physics Department, Weizmann Institute of Science, Rehovot 7610001, Israel\\
$^2$Institute for Theoretical Physics, University of Amsterdam, Science Park 904, 1098 XH Amsterdam, The Netherlands\\
$^3$School of Engineering and Applied Sciences, Harvard University, Cambridge 02138, USA}

\begin{abstract}
Identifying heterogeneous structures in glasses --- such as localized soft spots --- and understanding structure-dynamics relations in these systems remain major scientific challenges. Here we derive an exact expression for the local thermal energy of interacting particles (the mean local potential energy change due to thermal fluctuations) in glassy systems by a systematic low-temperature expansion. We show that the local thermal energy can attain anomalously large values, inversely related to the degree of softness of localized structures in a glass, determined by a coupling between internal stresses --- an intrinsic signature of glassy frustration ---, anharmonicity and low-frequency vibrational modes. These anomalously large values follow a fat-tailed distribution, with a universal exponent related to the recently observed universal $\omega^4$ density of states of quasi-localized low-frequency vibrational modes. When the spatial thermal energy field --- a `softness field' --- is considered, this power-law tail manifests itself by highly localized spots which are significantly softer than their surroundings. These soft spots are shown to be susceptible to plastic rearrangements under external driving forces, having predictive powers that surpass those of the normal-modes-based approach. These results offer a general, system/model-independent, physical-observable-based approach to identify structural properties of quiescent glasses and to relate them to glassy dynamics.
\end{abstract}

\maketitle

Understanding the glassy state of matter remains one of the greatest challenges in condensed-matter physics and materials science~\cite{Alexander1998, Dyre2006, Cavagna2009, Berthier2011, binder_kob_book}. In large part, this is due to the absence of well-established tools and concepts to quantify the disordered structures characterizing glassy materials --- in sharp contrast to their ordered crystalline counterparts --- and due to the lack of understanding of the relations between glassy structures and dynamics. Over the years, many attempts have been made to identify physical quantities that can indicate underlying local structures within glassy materials \cite{Glen,aharonov2007,Jack2014,Royall20151}. These indicators include, among others, free-volume~\cite{spaepen1977, spaepen2006,WidmerCooper2006}, internal stresses~\cite{tau-defects}, local elastic moduli~\cite{Barrat_2009}, local Debye-Waller factor~\cite{Asaph}, coarse-grained energy and density~\cite{Matharoo2006,Berthier2007}, locally favored structures~\cite{Coslovich2007, RoyallTanaka2008, Malins2013}, short- and medium-range order~\cite{Shi2005, Tanaka2005, Kawasaki2007} and various weighted sums over a system-dependent number of low-frequency normal modes~\cite{widmer2008irreversible, tanguy2010,manning2011,rottler_normal_modes,Mosayebi2014,Schoenholz2014, Ding2014}.

These quantities measure some properties of quiescent glasses, evaluated at or in the near vicinity of a mechanically (meta)stable state of a glass (an inherent structure). Some of these indicators are purely structural in nature, i.e.~they are obtained from the knowledge of particle positions alone, while others require in addition the knowledge of inter-particle interactions. Recently, the local yield stress --- the minimal local stress needed to trigger an irreversible plastic rearrangement --- has been proposed as a structural indicator~\cite{falk_local_yield}. It requires, however, to externally drive each local region in a glass to its nonlinear rearrangement threshold and hence belongs to a different class of structural indicators compared to those previously mentioned. The utility of each of the proposed indicators is usually assessed by looking for correlations between the revealed structures --- typically localized soft spots --- and glassy dynamics, either thermally-activated relaxation in the absence of external driving forces or localized irreversible plastic rearrangements under the application of global driving forces. In fact, a recent study established such structure-dynamics correlations by machine-learning techniques, leaving the precise physical nature of the underlying structural indicator unspecified~\cite{Cubuk2015, machine_learning}. These machine-learning-based structural indicators also belong to a different class of structural indicators since the training stage of the machine-learning algorithm requires knowledge of the plastic rearrangements themselves.

Some of the previously proposed structural indicators have revealed a certain degree of correlation between identified soft spots and dynamics, providing important evidence that pre-existing localized structures in a glass significantly affect its dynamics. Yet, oftentimes the physical foundations of the structural indicators remain unclear, and they are sometimes defined algorithmically, but not derived from well-established physical observables. Moreover, their statistical properties are not commonly addressed, the relations between them and other basic physical quantities are not established and the fundamental reasons for them being particularly sensitive to underlying heterogeneous structures in glasses remain elusive.

Here we propose a structural indicator of glassy `softness' --- the local thermal energy (LTE) --- which is a transparent \emph{physical observable} derived by a systematic low-temperature expansion. We use the exact expression for the LTE of interacting particles to elucidate the underlying physical factors --- most notably internal stresses, anharmonicity and nonlinear coupling to low-frequency vibrational modes --- that give rise to significant spatial heterogeneities of softness. We show that the LTE can attain anomalously large values, directly related to particularly soft regions in a glass, which follow a fat-tailed distribution. The power-law exponent characterizing this distribution is shown to be \emph{universal} and directly related to the recently observed universal $\omega^4$ density of states of quasi-localized low-frequency vibrational modes~\cite{modes_prl, Mizuno_arXiv}, constituting a link to a fundamental universal property of glassy systems. The LTE field, a `softness field', thus exhibits highly localized spots which are significantly softer than their surroundings. These soft spots are shown to be particularly susceptible to plastic rearrangements when the glass is being driven by external forces, having predictive powers that surpass those of the normal-modes-based approach~\cite{widmer2008irreversible, tanguy2010, manning2011,rottler_normal_modes}. As such, they can be identified with the long sought for glassy `flow defects', the so-called Shear-Transformation-Zones (STZ)~\cite{argon_st, falk_langer_stz}.

\section*{Physical Observables in the Low-temperature Limit}

Our starting point is the idea that the thermal average of local physical observables in a system equilibrated at a low temperature $T$ is expected to be sensitive to the system's underlying structure~\cite{Yohai}. Therefore, we first aim at deriving an expression for the thermal average of a general physical observable ${\cal A}$, $\langle {\cal A} \rangle_{_T}$, in the low-temperature limit. The latter is given by $\langle{\cal A}\rangle_{_T}\!=\!{\C Z}(T)^{-1}\!\int\!{\cal A}({\B x})\exp\!\left(\!{-\frac{{\C U}({\B x})}{k_B T}}\!\right) d{\B x}$, where the components of the vector ${\B x}$ represent the deviations of the system's degrees of freedom from a (possibly local) minimum of its energy $\C U({\B x})$, ${\C Z}(T)\!=\!\int\!\exp\!\left(\!{-\frac{{\C U}({\B x})}{k_B T}}\!\right) d{\B x}$ is the partition function and $k_B$ is the Boltzmann constant. $\langle{\cal A}\rangle_{_T}$ can be systematically expanded to leading order in $T$, yielding (see {\color{blue}{\em Supporting Information}})
\begin{equation}
 \label{eq:expansion}
\frac{\langle {\cal A} \rangle_{_T}-{\cal A}^{(0)}}{\tfrac{1}{2}k_B T}\!\simeq\! \frac{\partial^2\!{\cal A}}{\partial {\B x}\partial {\B x}}\!:\!\calBold{M}^{-1} - \frac{\partial {\cal A}}{\partial {\B x}}\!\cdot\!\calBold{M}^{-1}\!\cdot\calBold{U}'''\!:\!\calBold{M}^{-1} \ ,
\end{equation}
where $\calBold{M}\!\equiv\!\frac{\pa^2 {\C U}}{\pa{\B x}\pa{\B x}}$ is the dynamical matrix, $\calBold{U}'''\!\equiv\!\frac{\partial^3 {\C U}}{\partial {\B x} \partial {\B x} \partial {\B x}}$ is a third-order anharmonicity tensor and $\displaystyle {\cal A}^{(0)}\!\equiv\!\textstyle\lim_{T\to 0}\langle {\cal A} \rangle_{_T}$. All derivatives are evaluated at the minimum of $\C U$, i.e. at ${\B x}\!=\!{\mathBold{0}}$. In obtaining~\eqref{eq:expansion}, higher order terms in $T$ were neglected. In the $T\!\to\!0$ limit, these terms vanish and the right-hand-side (RHS) of~\eqref{eq:expansion} represents an intrinsic property of an inherent structure, independent of temperature.

To gain some understanding of the physics encapsulated in~\eqref{eq:expansion}, let us briefly consider a few physical observables. Consider first the total energy ${\cal A}\=\C U({\B x})$ in the quadratic (harmonic) approximation. In this case, the first (harmonic) term on the RHS of~\eqref{eq:expansion} equals the number of degrees of freedom $N$ and the second (anharmonic) term vanishes due to mechanical equilibrium, $\frac{\partial {\C U}}{\partial {\B x}}\={\mathBold{0}}$. Consequently, we obtain $\langle {\C U}\rangle_{_T}\!-{\C U}^{(0)}\!=\!\tfrac{1}{2}N k_B T$, which is nothing but the equipartition theorem in the harmonic approximation~\cite{chaikin2000principles}. Consider then a system whose energy ${\C U}(X)$ depends on a single (scalar) macroscopic degree of freedom $X$, representing changes in its linear dimension relative to a reference stable state ${X}\!=\!0$. In this case, the first (harmonic) term on the RHS of~\eqref{eq:expansion} vanishes and we obtain $\langle {X} \rangle_{_T}\!\simeq\!-\tfrac{1}{2}\,{\C U}'''({\C U}'')^{-2} k_B T\!+\!{\C O}(T^2)$, where a prime denotes a derivative with respect to $X$. This describes linear thermal expansion, which is well-known to be an intrinsically anharmonic physical effect proportional to ${\C U}'''$~\cite{chaikin2000principles}. These examples both show that~\eqref{eq:expansion} is fully consistent with well-established results (equipartition and thermal expansion) and highlight the anharmonic nature of the second term on the RHS of~\eqref{eq:expansion}.

\begin{figure*}[ht]
\centering
\begin{tabular}{ccc}
\includegraphics[width=0.97\linewidth]{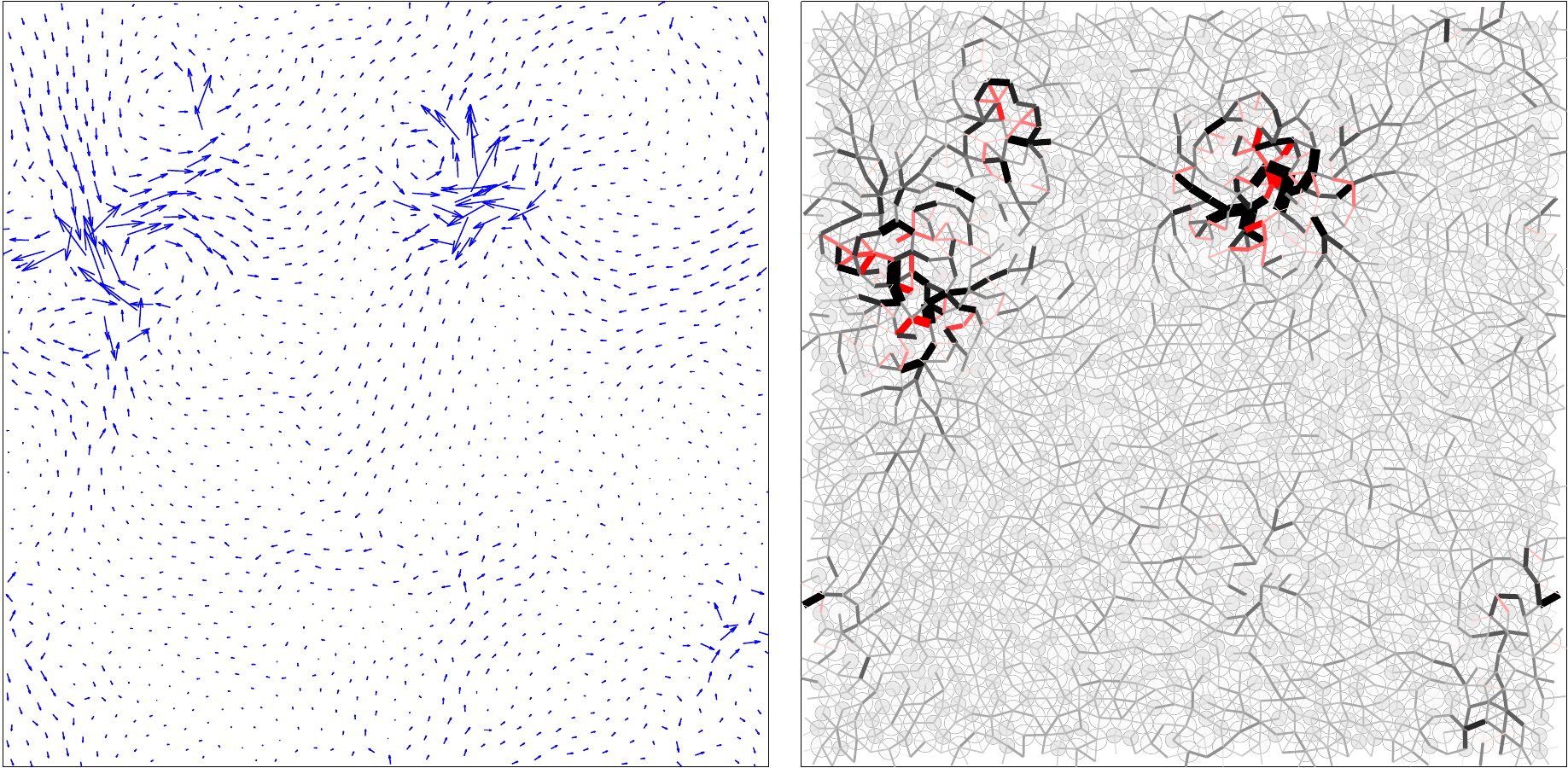}
\end{tabular}
\caption{(left) The normalized thermal displacement vector $\calBold{X}$, defined in~\eqref{eq:thermal displacements}, for a 2DIPL system (see text) of $N\!=\!1600$. (right) A spatial map of the normalized LTE ${\C E}_\alpha$, defined in~\eqref{eq:thermal energy}, for the same glass realization shown in the left panel. Line thickness and opacity represent the LTE, with red (black) representing negative (positive) LTE.}
\label{fig:thermal_displacements}
\end{figure*}

The examples presented above focused on macroscopic (global) scalar observables. As our main interest is in spatial heterogeneity, we consider now microscopic (local) observables defined at the particles' level. We thus focus on the microscopic generalization of $\langle{X}\rangle_{_T}$: the thermal displacement vector $\langle{\B x}\rangle_{_T}$, which represents the variation of the mean positions of particles about the equilibrium state once thermal fluctuations are introduced. Using~\eqref{eq:expansion}, the normalized thermal average of $\B x$ in the $T\!\to\!0$ limit takes the form
\begin{equation}
\calBold{X} \equiv \lim_{T\to 0}\frac{\langle {\B x} \rangle_{_T}}{\tfrac{1}{2}k_B T} = -\calBold{M}^{-1}\!\cdot\calBold{U}'''\!:\!\calBold{M}^{-1}\ .\label{eq:thermal displacements}
\end{equation}
Note the analogy between~\eqref{eq:thermal displacements} --- which features a quadratic (nonlinear) coupling between the anharmonicity tensor $\calBold{U}'''$ and the inverse of the dynamical matrix $\calBold{M}^{-1}$ --- and the expression given above for $\langle {X} \rangle_{_T}$. The components of the normalized thermal displacement vector ${\C X}_i$ in~\eqref{eq:thermal displacements} should be distinguished from the local Debye-Waller factor $x^2_i$~\cite{Asaph}, whose thermal average according to~\eqref{eq:expansion} is given by $\langle x^2_i \rangle_{_T}\=({\C M}^{-1})_{ii}\,k_B T$ (no summation is implied). While $\langle x^2_i \rangle_{_T}$ is completely given by the first term on RHS of~\eqref{eq:expansion}, which involves a single contraction of the inverse of the dynamical matrix $\calBold{M}^{-1}$, ${\C X}_i$ is completely given by the second term, which involves two contractions with $\calBold{M}^{-1}$. As will be shown below, this distinction makes a qualitative difference. Moreover, ${\C X}_i$ is directly sensitive to anharmonicity, while $\langle x^2_i \rangle_{_T}$ is independent of it. $\calBold{X}$, plotted in Fig.~\ref{fig:thermal_displacements} for a 2D model glass, is shown to exhibit significant spatial heterogeneity, suggesting that it is particularly sensitive to localized soft structures in glasses.

\section*{Local Thermal Energy}

The normalized thermal displacement vector $\calBold{X}$, defined in~\eqref{eq:thermal displacements} and shown to exhibit strong spatial heterogeneity in Fig.~\ref{fig:thermal_displacements}, contributes to the thermal average of {\em any} physical observable $\langle {\cal A} \rangle_{_T}$ that features $\frac{\partial {\cal A}}{\partial {\B x}}\!\ne\!{\mathBold{0}}$ at ${\B x}\={\mathBold{0}}$. It is important to emphasize the counter-intuitive result that for observables with $\frac{\partial {\cal A}}{\partial {\B x}}\!\ne\!{\mathBold{0}}$, anharmonicity appears to be important at vanishingly small temperatures, independent of how well the harmonic approximation for the energy holds. Thus, on the face of it, the normalized thermal displacements $\calBold{X}$ could have been a good candidate for an indicator of `softness' of the underlying structure. However, we aim at proposing an observable that naturally `filters out' the regions of homogeneous, collective-translation-like, motion exhibited by the thermal displacements, further exposing localized soft structures that exhibit large gradients.

Our goal now is to identify a physical observable $\C A$ that can potentially serve as a `softness field', i.e.~a local scalar that features a nonvanishing first spatial derivative and is particularly sensitive to gradients of $\calBold{X}$. Inspired by~\cite{Yohai}, an observable that naturally suggests itself is the local potential energy $\varepsilon_\alpha$, where $\alpha$ represents any pair of interacting particles and ${\C U}\=\sum_\alpha\! \varepsilon_\alpha$. Using Eqs.~(\ref{eq:expansion})-(\ref{eq:thermal displacements}), we then define
\begin{equation}
{\C E}_\alpha \equiv \lim_{T \to 0}\!\frac{\langle{\varepsilon}_\alpha\rangle_{_T}-{\varepsilon}_\alpha^{(0)}}{\tfrac{1}{2}k_B T} = \frac{\pa {\B f}_\alpha}{\partial \B x}\!:\!\calBold{M}^{-1} + {\B f}_\alpha \!\cdot\!  \calBold{X} , \label{eq:thermal energy}
\end{equation}
where ${\B f}_\alpha \equiv \frac{\pa \varepsilon_\alpha}{\pa \B x}$ is the internal force vector acting between particles defining the interaction $\alpha$.

Mechanical equilibrium at particle $i$ implies that the \textit{sum} of all the forces acting on it vanishes. In systems with no internal frustration, internal forces/stresses do not exist and this sum is trivially satisfied by having ${\B f}_{\alpha}\=\mathBold{0}$ for all $\alpha$'s. In systems with internal frustration, however, internal forces/stresses generically emerge, ${\B f}_{\alpha}\!\ne\!\mathBold{0}$. In the former case, the second term on the RHS of~\eqref{eq:thermal energy} vanishes. Such internal-stress-free disordered systems were studied in~\cite{Yohai}, where it was shown that under these conditions ${\C E}_\alpha$ is universally bounded between $0$ and $1$. This suggests that significant spatial heterogeneity in ${\C E}_\alpha$ cannot emerge in internal-stress-free systems.

An intrinsic signature of glassy systems is the existence of internal frustration~\cite{Tarjus2005} that leads to the emergence of internal forces/stresses, ${\B f}_{\alpha}\!\ne\!\mathBold{0}$~\cite{Alexander1998}. Consequently we expect the ${\B f}_\alpha\!\cdot\!\calBold{X}$ term on the RHS of~\eqref{eq:thermal energy} to be generically non-zero for glasses. As $\calBold{X}$ is already known to exhibit strongly localized structures, cf.~Fig.~\ref{fig:thermal_displacements} (left), we expect ${\B f}_\alpha\!\cdot\!\calBold{X}$ to expose localized regions with a very large concentration of the normalized LTE ${\C E}_\alpha$. In fact, we expect the scalar product of ${\B f}_\alpha$ with $\calBold{X}$ to amplify the spatial heterogeneity in $\calBold{X}$. To understand this, note that ${\B f}_\alpha$ is actually a force-dipole composed of two forces acting along the line connecting the particles that define the interaction $\alpha$, in {\em opposite} directions. Therefore, ${\B f}_\alpha\!\cdot\!\calBold{X}$ is {\em exactly} the difference between the values of $\calBold{X}$ at the positions of the particles defining the interaction $\alpha$, projected along the line connecting them, multiplied by $|{\B f}_\alpha|$. Consequently, regions of homogeneous thermal displacements are expected to feature small values of ${\B f}_\alpha\!\cdot\!\calBold{X}$, while heterogeneous regions --- cf.~Fig.~\ref{fig:thermal_displacements} (left) --- are expected to feature much larger values.

\begin{figure}[ht]
\centering
\begin{tabular}{ccc}
\includegraphics[width=0.98\linewidth]{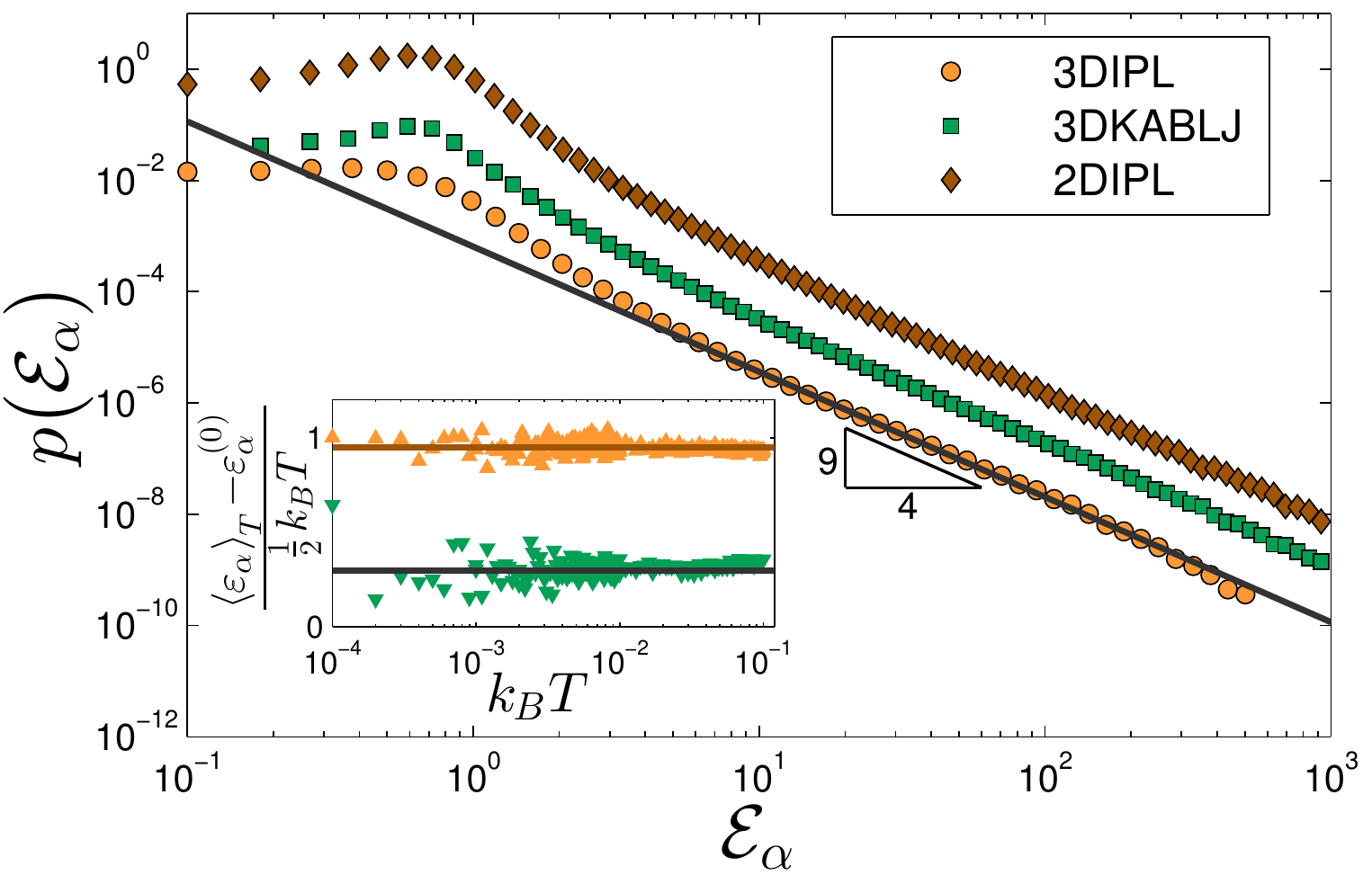}
\end{tabular}
\caption{Distributions of LTE --- $p({\C E}_\alpha)$ --- measured for three model glasses in 2D and 3D (see text for details), shifted vertically for visibility. We find a universal form $p({\C E}_\alpha) \sim {\C E}_\alpha^{-9/4}$ at large LTEs, independent of model or spatial dimension. Inset: Molecular dynamics validation of Eq.~(\ref{eq:thermal energy}) for two random interactions in a model glass. The continuous lines represent the exact expression for ${\C E}_\alpha$.}
\label{fig:thermal_energy}
\end{figure}

To test these ideas, we plot in~Fig.~\ref{fig:thermal_displacements} (right) the normalized LTE ${\C E}_\alpha$ for the same glass realization shown in the left panel. The result is striking: ${\C E}_\alpha$ attains anomalously large values (both positive and negative) in localized regions where $\calBold{X}$ exhibits marked heterogeneity. This observation provides strong visual evidence, to be quantified below, that ${\C E}_\alpha$ can be used to define a `softness field' that clearly identifies localized soft spots in glasses. Finally, note that ${\C E}_\alpha$ can be also measured directly by tracking thermal fluctuations in low $T$ dynamics. Two examples obtained by finite $T$ Molecular Dynamics (MD) simulations are shown in Fig.~\ref{fig:thermal_energy} (inset), demonstrating perfect agreement with the exact expression in~\eqref{eq:thermal energy}.

\section*{Universal Anomalous Statistics}

To quantify the degree of `softness' of soft spots revealed by ${\C E}_\alpha$ --- cf.~Fig.~\ref{fig:thermal_displacements} (right) --- and its probability of occurrence, we focus next on the statistical properties of ${\C E}_\alpha$. To this aim, we argue that the statistics of normalized thermal energies ${\C E}_\alpha$ can be related to the density of vibrational frequencies $D(\omega)$. In particular, the form of Eqs.~(\ref{eq:thermal displacements})-(\ref{eq:thermal energy}) suggests that soft vibrational modes, i.e.~modes with small frequencies $\omega$, give rise to large values of ${\C E}_\alpha$ due to the appearance of the inverse of the dynamical matrix $\calBold{M}^{-1}$. Recently, it has been observed that low-frequency vibrations in glassy materials appear in two qualitatively different species, one is ordinary long-wavelength plane-waves and the other is disorder-induced soft glassy modes. The former are spatially extended objects, while the latter are quasi-localized objects characterized by a disordered core and a power-law tail~\cite{modes_prl}. Moreover, long-wavelength plane-waves follow a Debye density of states (DOS) $D_D(\omega)\!\sim\!\omega^{\dbar-1}$, in $\dbar$ dimensions, while soft glassy modes follow a universal DOS $D_G(\omega)\!\sim\!\omega^{4}$~\cite{modes_prl, Mizuno_arXiv}. We stress that our focus here is on generic glasses, which do not dwell near a jamming transition, where the physics is expected to change.

To proceed, note that ${\C E}_\alpha$ in~\eqref{eq:thermal energy} has one contribution that involves a single contraction with $\calBold{M}^{-1}$ and another one that involves two contractions with $\calBold{M}^{-1}$, therefore the latter is expected to dominate the former. Consequently, we write ${\C E}_\alpha\!\sim\! {\B f}_\alpha \!\cdot\! \calBold{X}$ whose eigen-decomposition takes the form
\begin{equation}
{\C E}_\alpha\sim\sum_{i,j} \frac{({\B f}_\alpha\!\!\cdot\!\mathBold{\Psi}_i)\,c_{ijj}}{\omega_i^{2}\,\omega_j^{2}}\,\,\quad\hbox{with}\,\,\quad c_{ijj}\equiv\calBold{U}'''\tripleCdot\mathBold{\Psi}_i\mathBold{\Psi}_j\mathBold{\Psi}_j \ ,
\label{eq:eigendecomposition}
\end{equation}
where $i,j$ run over all of the vibrational modes $\mathBold{\Psi}_i$, defined by the eigenvalue equation $\calBold{M} \!\cdot\!\mathBold{\Psi}_i\=\omega^2_i\,\mathBold{\Psi}_i$.

We argue that low-frequency plane-waves and quasi-localized soft glassy modes make qualitatively different contributions to the double sum in~\eqref{eq:eigendecomposition}. To see this, note that similarly to the discussion about the dipolar nature of ${\B f}_\alpha$ above, each contraction of $\calBold{U}'''$ with a vibrational mode is proportional to the mode's spatial derivative (cf.~Fig.~3 in~\cite{micromechanics}). For low-frequency plane-waves, each such derivative is proportional to the frequency $\omega$, while for quasi-localized soft glassy modes the derivative is expected to attain a characteristic value that is nearly independent of frequency. Consequently, since $c_{ijj}\!\sim\!\omega^3$ and ${\B f}_\alpha\!\!\cdot\!\mathBold{\Psi}_i\!\sim\!\omega$ for plane-waves (which we have numerically verified), we expect their contribution to be negligible compared to that of quasi-localized soft glassy modes, and hence the above double sum is now understood to be dominated by the latter. Next, since different quasi-localized soft glassy modes are spatially well separated, we expect $c_{ijj}$ for $i\!\ne\!j$ to be much smaller than $c_{iii}$ such that ${\C E}_\alpha\!\sim\!\sum_{i} ({\B f}_\alpha\!\cdot\!\mathBold{\Psi}_i)\,c_{iii}\,\omega_i^{-4}$. Finally, as the internal force ${\B f}_\alpha$ is localized at the $\alpha$-th interaction, only the glassy mode that is localized there will contribute to the sum, leading to
\begin{equation}
{\C E}_\alpha \sim \omega^{-4} \ .
\label{eq:scaling1}
\end{equation}
Equation~(\ref{eq:scaling1}), which is verified below, establishes an important relation between the LTE ${\C E}_\alpha$ and the frequency of vibrational modes $\omega$. In fact, it constitutes a relation between ${\C E}_\alpha$ and the local stiffness $\kappa\!\equiv\!\omega^2$, ${\C E}_\alpha\!\sim\!\kappa^{-2}$, showing that particularly soft excitations, $\kappa\!\to\!0$, correspond to anomalously large values of the LTE ${\C E}_\alpha$. This justifies the assertion that ${\C E}_\alpha$ quantifies the the degree of softness of glassy structures.

Using~\eqref{eq:scaling1} and the universal relation $D_G(\omega)\!\sim\!\omega^4$, the probability distribution function $p({\C E}_\alpha)$ is obtained as
\begin{equation}
p({\C E}_\alpha) = D_G\!\left[\omega({\C E}_\alpha)\right]\,\,\frac{d\omega({\C E}_\alpha)}{d{\C E}_\alpha} \,\sim\, {\C E}_\alpha^{-1}\,\,{\C E}_\alpha^{-5/4} \,\sim\, {\C E}_\alpha^{-9/4} \ . \label{eq:scaling2}
\end{equation}
Note that in the above discussion we implicitly used the fact that the magnitude of the internal forces $|{\B f}_\alpha|$ has a characteristic value, as shown in~{\color{blue}{\em Supporting Information}}. The prediction in~\eqref{eq:scaling2} has far reaching implications. First, it suggests that the physical observable ${\C E}_\alpha$, i.e.~the LTE, effectively filters out the effect of low-frequency plane-waves, which are known to obscure the origin of many glassy effects~\cite{micromechanics,manning2015,nonlinear_modes_scipost}. In fact, when low-frequency plane-waves coexist with quasi-localized soft glassy modes in the same frequency range, they hybridize such that glassy modes acquire spatially extended background displacements and appear to lose their quasi-localized nature. The derivation leading to~\eqref{eq:scaling2} assumed that ${\C E}_\alpha$ is insensitive to hybridization and that the $D_G(\omega)\!\sim\!\omega^4$ distribution remains physically meaningful --- i.e.~it still characterizes the probability to find a soft localized structure in a glass --- even in the presence of hybridization, when it cannot be directly probed by a harmonic normal modes analysis. Second, the prediction in~\eqref{eq:scaling2} rationalizes the existence of anomalously soft localized spots in glassy materials and predicts its probability.

To test the prediction in~\eqref{eq:scaling2} and its degree of universality, we performed extensive numerical simulations of different computer glass-forming models: (i) a binary system of point-like particles interacting via inverse power-law purely repulsive pairwise potentials in 2D (2DIPL) and 3D (3DIPL)~\cite{modes_prl}; (ii) the canonical
Kob-Andersen binary Lennard-Jones (3DKABLJ) system~\cite{kablj} in 3D (see~{\color{blue}{\em Supporting Information}} for details about models and methods), in order to extract the statistics of ${\C E}_\alpha$ according to~\eqref{eq:thermal energy}. The results are summarized in Fig.~\ref{fig:thermal_energy}. All of the glasses considered exhibit a power-law tail with a universal exponent fully consistent with the theoretically predicted $-9/4$ exponent. These results lend strong support to the prediction in~\eqref{eq:scaling2} and therefore also implicitly to its underlying assumptions. The results presented in this section explain the physical origin of the sensitivity of ${\C E}_\alpha$ to soft glassy structures, elucidate its anomalous statistical properties and establish a relation between its statistical properties and the recently observed universal $\omega^4$ density of states of quasi-localized low-frequency vibrational modes~\cite{modes_prl}, a fundamental property of glasses. Next, we would like to explore the possibility of defining mesoscopic soft spots based on ${\C E}_\alpha$ and the their predictive powers.

\section*{Softness Field and Predicting Plastic Rearrangements}

\begin{figure}[ht]
\centering
\includegraphics[width=1.0\linewidth]{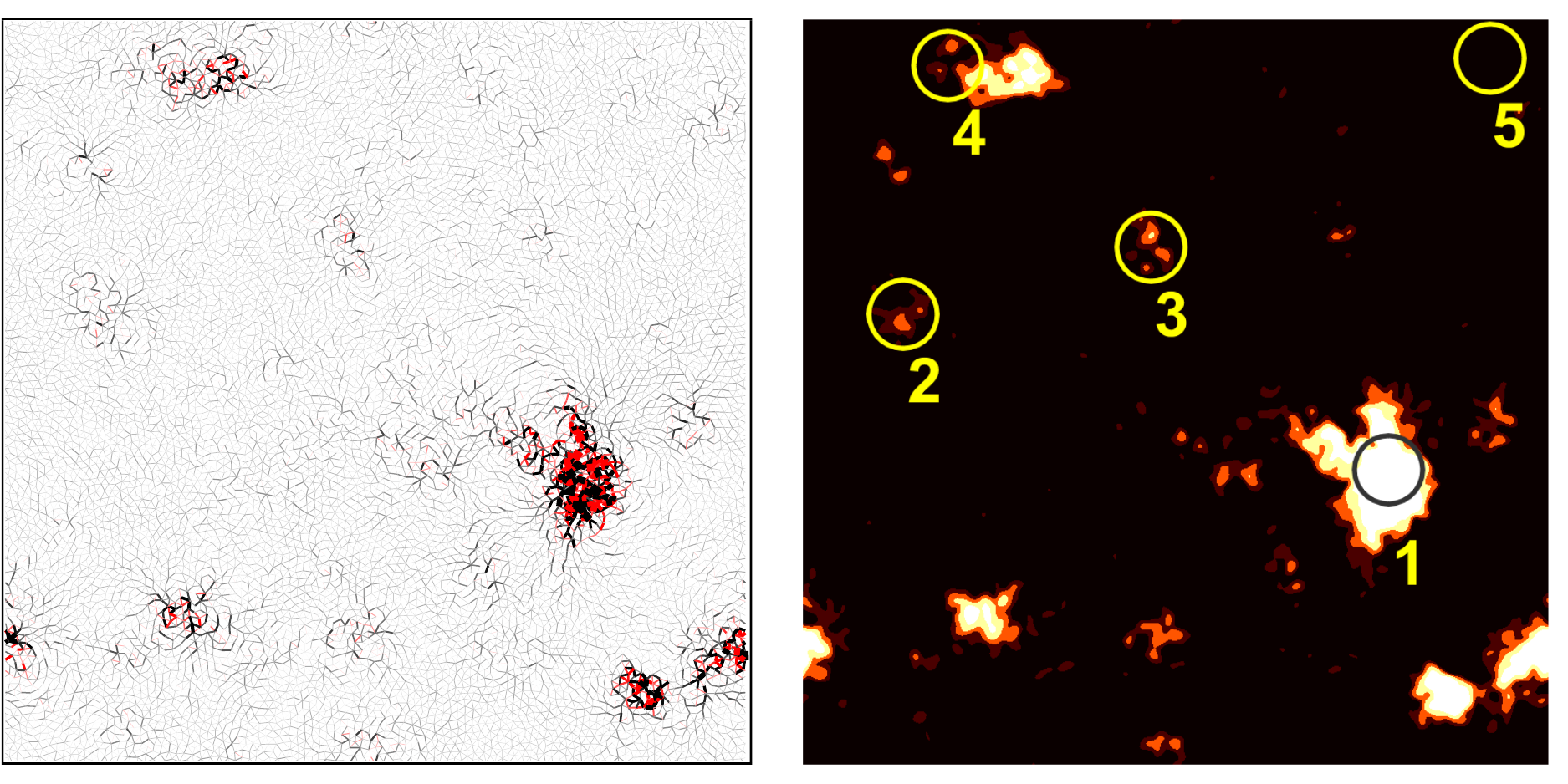}
\caption{(left) LTE field in a 2DIPL system of $N\!=\!10000$, same as in Fig.~\ref{fig:thermal_displacements}. (right) coarse-grained softness field, see text. Enumerated by occurrence order are the loci of plastic instabilities that occur upon application of quasistatic shear deformation.}
\label{fig:predictability}
\end{figure}

The normalized LTE ${\C E}_\alpha$ is microscopically defined for any interaction $\alpha$. In Fig.~\ref{fig:predictability} (left) we present yet another example of the spatial map of ${\C E}_\alpha$, here for a larger system compared to Fig.~\ref{fig:thermal_displacements} (right). A continuous field can be naturally constructed by coarse-graining $|{\C E}_\alpha|$ on a scale larger than the particles scale. We use $|{\C E}_\alpha|$ because anomalously large negative and positive values of ${\C E}_\alpha$ are strongly correlated in space. Coarse-graining is achieved by discretizing space into bins containing at least two bonds each, assigning a bin with softness obtained by averaging the values of $|{\C E}_\alpha|$ of bonds belonging to it and finally by averaging the bin's value with the values of all bins in the first layer of neighboring bins (see~{\color{blue}{\em Supporting Information}}). Applying this procedure to Fig.~\ref{fig:predictability} (left) yields Fig.~\ref{fig:predictability} (right), which we treat as a `softness field'. Our goal now is to test the predictive powers of this softness field in relation to glassy dynamics. The latter, either thermally-activated relaxation in non-driven conditions or plastic rearrangements under external driving forces, entail crossing some activation barriers. Activation barriers revealed by soft localized vibrational modes $\mathBold{\Psi}_i$ of frequency $\omega_i$ are small, of order $\omega_i^6/c_{iii}^2$ in the leading anharmonic expansion of the energy~\cite{luka}. Hence, we expect that regions that feature large values of $|{\C E}_\alpha|$ will be particularly susceptible to plastic rearrangements.

To test this, we applied global quasi-static shear deformation in a certain direction, under athermal conditions, to each glass realization --- such as the one shown in Fig.~\ref{fig:predictability} (right) --- and measured the locations of the first few discrete irreversible plastic rearrangements, as described in~{\color{blue}{\em Supporting Information}}. The advantage of this $T\=0$ protocol is that it allows to uniquely and unquestionably identify the discrete irreversible plastic rearrangements. The locations of the first $5$ discrete irreversible plastic rearrangements (events) were superimposed on the softness field in Fig.~\ref{fig:predictability} (right). The first $4$ plastic events overlap soft spots identified by the softness field, indicating a high degree of predictiveness of ${\C E}_\alpha$.

To quantify the degree of predictiveness of the LTE ${\C E}_\alpha$, we extracted the location of soft spots from the spatial distribution of ${\C E}_\alpha$, for example the one shown in Fig.~\ref{fig:predictability}, as described in~{\color{blue}{\em Supporting Information}}. In addition to its location, each soft spot is characterized by its degree of softness, representing the average value of $|{\C E}_\alpha|$ in its near vicinity (see~{\color{blue}{\em Supporting Information}}). As the fat-tailed distribution in~\eqref{eq:scaling2} predicts very large variability in the degree of softness of different soft spots within a single glass realization and among different realizations, we define $\Delta_{\C E}$ of each soft spot as the maximal degree of softness in a given realization divided by the spot's degree of softness. That way we standardize the degree of softness such that the softest spot in each realization has $\Delta_{\C E}\=1$ and not-as-soft spots have $\Delta_{\C E}\!>\!1$. Then each plastic event of ordinal number $n$ ($n\=1$ for the first event, $n\=2$ for the second etc.) is associated with the soft spot that is closest to it in space (see~{\color{blue}{\em Supporting Information}}). We stress that the soft spots are extracted for the non-sheared system, and are not updated between plastic events.

\begin{figure}
\centering
\includegraphics[width=\linewidth]{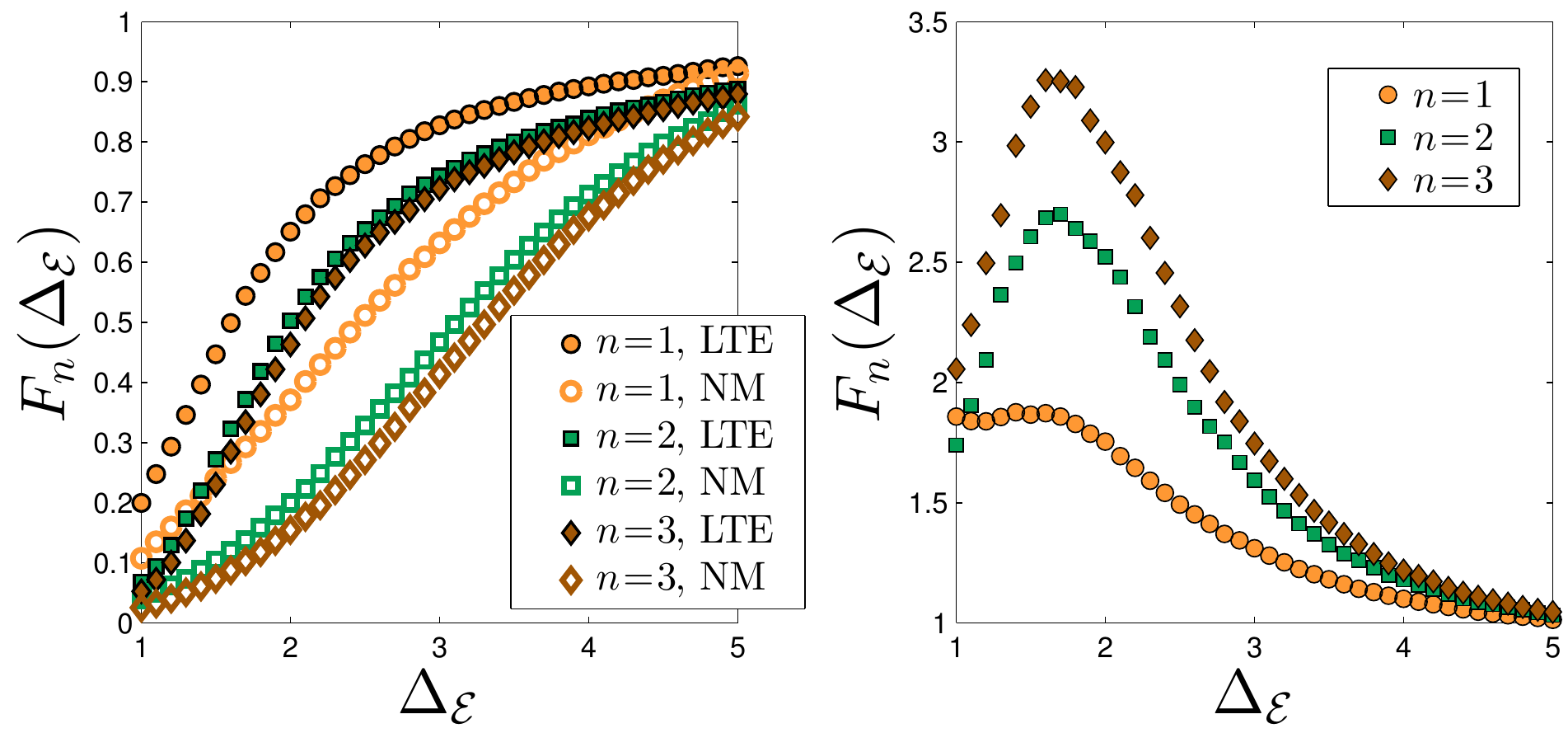}
\caption{(left) The cumulative distribution function $F_n(\Delta_{\C E})$, quantifying the fraction of plastic events of ordinal number $n\!=\!1, 2, 3$ being closest to soft spots characterized by a value equal or smaller than $\Delta_{\C E}$ (full symbols). The corresponding results based on the normal-modes (NM) approach (see text for details) are superimposed (empty symbols). (right) The ratio of $F_n(\Delta_{\C E})$ for the two approaches, $\delta\!F_n(\Delta_{\C E})$, is plotted for $n\!=\!1, 2, 3$. It is clearly observed that the thermal-energy-based approach significantly outperforms the normal-modes-based approach.}
\label{fig:comparison}
\end{figure}

The cumulative distribution function $F_n(\Delta_{\C E})$, quantifying the fraction of plastic events of ordinal number $n$ being closest to soft spots characterized by a value equal or smaller than $\Delta_{\C E}$, is constructed by collecting data from $5000$ independent simulations of 2DIPL computer glasses. $F_n(\Delta_{\C E})$ for $n\=1, 2, 3$ is shown in Fig.~\ref{fig:comparison} (left, full symbols). As expected, the smaller $n$ the larger the predictive power. Moreover, it is observed that about $20\%$ of the first plastic events (i.e.~$n\=1$) are predicted by the softest spot in each realization and nearly $70\%$ are predicted by soft spots with $\Delta_{\C E}\!\le\!2$. To assess how good these predictive powers are, we need some reference case to compare to, which we consider next.

\subsection*{Comparison to the normal-modes-based approach} Among the many structural indicators studied over the years, cf.~the introduction above, the normal-modes-based approach~\cite{widmer2008irreversible, tanguy2010, manning2011,rottler_normal_modes} stands out according to the relatively high correlations between structure and dynamics it exhibits. The basic idea behind this approach is that while a single low-lying normal mode $\mathBold{\Psi}_i$ does not clearly exhibit localized structures, possibly due to hybridization, some weighted sum over a system-dependent number of normal modes does reveal such structures. We use this approach here in order to compare its predictions to the predictions obtained above based on the LTE. In particular, we follow~\cite{Asaph} and construct maps analogous to Fig.~\ref{fig:predictability} (left) and Fig.~\ref{fig:thermal_displacements} (right) by summing the norm squared of the components of low-lying normal modes $\mathBold{\Psi}_i$ at each particle over the first $30$ non-zero modes, i.e.~$\sum_{i=1}^{30}|\mathBold{\Psi}^{(j)}_{i}|^2$ for every particle $j$. Here $\mathBold{\Psi}^{(j)}_{i}\!\equiv\!(\Psi^{(j)}_{i,x}, \Psi^{(j)}_{i,y})$ are the components of the normal mode $\mathBold{\Psi}_i$ at particle $j$ and $x, y$ are the axes directions in a global 2D Cartesian coordinate system.

Once the normal modes maps are constructed (see~{\color{blue}{\em Supporting Information}} for more details), we apply to them the {\em same} procedure described above and calculated the cumulative distribution function $F_n(\Delta_{\C E})$ based on them. The results are superimposed on the LTE results in Fig.~\ref{fig:comparison} (left, empty symbols). The comparison reveals that the thermal-energy-based approach significantly outperforms the normal-modes-based approach. This is quantified in Fig.~\ref{fig:comparison} (right), where we plot the ratio of $F_n(\Delta_{\C E})$ for the two approaches for $n\=1, 2, 3$, $\delta\!F_n(\Delta_{\C E})$, demonstrating that the thermal-energy-based approach outperforms the normal-modes-based approach by up to a factor of $1.85$ for $n\=1$ and up to a factor of $3.3$ for $n\=3$.

We thus conclude that the LTE has predictive powers that surpass those of the normal-modes-based approach. Can we also assess its predictive powers in absolute terms? To address this question, one should note that soft spots are expected to be anisotropic objects~\cite{lemaitre2006_avalanches,micromechanics} characterized by orientation and polarity, and hence feature variable coupling to shearing in various directions. That is, they are expected to be spin-like objects. Consequently, a spot which is very soft in a given direction may not undergo a rearrangement if the projection of the driving force on its soft direction is small. Hence, the optimal predictive power based on the degree of softness alone --- a scalar measure --- may be significantly smaller than unity. In particular, assuming a uniform/isotropic orientational distribution of equally-soft spots, a naive estimation indicates that only $25\%$ of them will rearrange under shearing in a given direction. As a result, the $\sim\!20\%$ predictive power of the softest soft in each realization, cf.~Fig.~\ref{fig:comparison} (left, full symbols, $n\=1$), may in fact be not so far from the optimal scalar predictiveness level. The optimal scalar predictiveness issue certainly deserves further investigation.

\section*{Conclusion}

We have shown that the low-temperature LTE ${\C E}_\alpha$ is a physical observable that is particularly sensitive to localized soft structures in glasses. ${\C E}_\alpha$ effectively filters out the contribution of long-wavelength plane-waves, hence it is dominated by soft glassy vibrational modes alone. This property allows to establish a quantitative relation between the recently observed universal distribution of soft glassy vibrational modes, $D_G(\omega)\!\sim\!\omega^{4}$ in the limit of small frequencies $\omega$, and the distribution of the LTE, $p({\C E}_\alpha)\!\sim\! {\C E}_\alpha^{-9/4}$ in the limit of large ${\C E}_\alpha$. This universal anomalous, fat-tailed distribution of ${\C E}_\alpha$ has been supported by extensive simulations on various computer glass-former in 2D and 3D.

While the problem of coexistence and hybridization of long-wavelength plane-waves and soft vibrational modes, which has hampered a direct observation of soft quasi-localized glassy modes and their statistical distribution for a long time, will be addressed elsewhere, we stress that our results have potentially important implications in this context. The universal fat-tailed distribution $p({\C E}_\alpha)\!\sim\!{\C E}_\alpha^{-9/4}$ has been theoretically derived based on the DOS of soft quasi-localized vibrational modes $D_G(\omega)\!\sim\!\omega^{4}$. Yet, the LTE ${\C E}_\alpha$ is a physical quantity that is defined without any explicit reference to soft quasi-localized vibrational modes or to any harmonic normal modes analysis. Consequently, it should be valid in the thermodynamic limit where the harmonic normal modes analysis may neither cleanly reveal soft quasi-localized vibrational modes nor their $\omega^{4}$ DOS. As such, it suggests that the $\omega^{4}$ distribution has a physical meaning that goes beyond the eigenvalues of harmonic normal modes, where $\kappa\=\omega^2$ is a generalized measure the stiffness of localized soft glassy structures~\cite{nonlinear_modes_scipost}.

The universal anomalous distribution of ${\C E}_\alpha$ and its relation to the universal localized glassy modes DOS imply the existence of highly localized and soft structures in glassy materials. Consequently, ${\C E}_\alpha$ forms a softness field that naturally reveals soft spots. These soft spots are expected to be characterized by particularly small activation barriers and hence to predict the loci of plastic rearrangements under shearing. As such, these soft spots are natural candidates for STZ~\cite{argon_st, falk_langer_stz}. The predictive powers of the LTE have been substantiated by extensive numerical simulations and have been shown to be superior to those of the normal-modes-based structural indicator.

Our approach offers a general, system/model-independent, physical-observable-based framework to identify structural properties of quiescent glasses and to relate them to glassy dynamics. In particular, the identified field of soft spots and its time-evolution under external driving forces should play a major role in theories of plasticity of amorphous materials, serving to define a population of STZ~\cite{falk_langer_stz, bouchbinder2007athermal, bouchbinder2009nonequilibrium, bouchbinder2011linear, falk2011deformation}. The predictive powers of our approach have been demonstrated here for plastic rearrangements in athermal quasi-statically driven systems. An important future challenge would be to test whether and to what extent these predictive powers persist at finite temperatures --- possibly up to the glass transition region --- and finite strain rates. It should also be tested against thermally-activated relaxation in the absence of external driving forces. Finally, as mentioned above, an interesting direction would be to go beyond the scalar degree of softness measure by incorporating orientational information into a generalized structural indicator.

{\bf Acknowledgement} E.B.~acknowledges support from the Harold Perlman Family Foundation, and the William Z. and Eda Bess Novick Young Scientist Fund.


\onecolumngrid
\newpage
\begin{center}
	\textbf{\large Supporting Information}
\end{center}
\twocolumngrid

\setcounter{equation}{0}
\setcounter{figure}{0}
\setcounter{section}{0}
\setcounter{table}{0}
\setcounter{page}{1}
\makeatletter

\newcommand{\be}{\begin{equation}}
\newcommand{\ee}{\end{equation}}
\newcommand{\ba}{\begin{eqnarray}}
\newcommand{\ea}{\end{eqnarray}}
\newcommand{\kb}{k_{\scriptscriptstyle\rm B}}
\newcommand{\wAM}{\omega_{\scriptscriptstyle\rm AM}}
\newcommand{\rhomax}{\rho_{\scriptscriptstyle\rm MAX}}
\newcommand{\phiRCP}{\phi_{\scriptscriptstyle\rm RCP}}
\newcommand{\phiC}{\phi_{\scriptscriptstyle\rm C}}
\newcommand{\phiR}{\phi_{\scriptscriptstyle\rm R}}
\newcommand{\phio}{\phi_{\scriptscriptstyle\rm 0}}
\newcommand{\omegaAM}{\omega_{\scriptscriptstyle\rm AM}}

\newcommand{\fm}{\langle f \rangle}
\newcommand{\hR}{h_{\scriptscriptstyle\rm R}}
\newcommand{\ho}{h_{\scriptscriptstyle\rm 0}}
\newcommand{\epsilonR}{\epsilon_{\scriptscriptstyle\rm R}}
\newcommand{\omegaBP}{\omega_{\scriptscriptstyle\rm BP}}

\newcommand{\eq}[1]{\begin{align}#1\end{align}}
\newcommand{\eqs}[1]{\begin{align*}#1\end{align*}}
\newcommand{\ffrac}[2]{\mbox{$\frac{#1}{#2}$}}
\def\half{\mbox{$\frac{1}{2}$}}
\def\quarter{\mbox{$\frac{1}{4}$}}
\def\eighth{\mbox{$\frac{1}{8}$}}
\def\halfs{\mbox{$1/2$}}
\def\twothirds{\mbox{$\frac{2}{3}$}}
\def\tr{\mbox{tr}}
\newcommand{\sigmab}{\bm{\hat{\sigma}}}
\newcommand{\p}{\partial}

\newcommand{\dz}{\delta z}
\newcommand{\PP}{\mathbb{P}}
\newcommand{\Gb}{\bm{G}}

\def\zbar{\overline{z}}
\def\eb{\bm{e}}
\def\rb{\bm{r}}
\def\qb{\bm{q}}
\def\fb{\bm{f}}
\def\Ub{\bm{U}}
\def\vb{\bm{v}}
\def\lb{\bm{\ell}}
\def\nb{\bm{n}}
\def\mb{\bm{m}}
\def\ij{\langle ij \rangle}
\def\kl{\langle kl \rangle}
\def\epsb{\bm{\hat{\varepsilon}}}
\def\delb{\bm{\hat{\delta}}}
\def\Fk{|F\rangle}
\def\Rk{|R\rangle}
\def\FDk{|F^D\rangle}
\def\fk{|f\rangle}
\def\ftk{|\tilde{f}\rangle}
\def\tauk{|\tau\rangle}
\def\Uk{|U\rangle}
\def\Ukf{|U^f \rangle}
\def\vk{|v\rangle}
\def\omegak{|\omega\rangle}
\def\vkf{|v^f\rangle}
\def\omegakf{|\omega^f \rangle}
\def\curlyS{\mathcal{S}}
\def\curlyE{\mathcal{E}}
\def\curlyU{\mathcal{U}}
\def\curlyM{\mathcal{M}}
\def\curlyP{\mathcal{P}}
\def\curlyR{\mathcal{R}}
\def\curlyT{\mathcal{T}}
\def\curlyN{\mathcal{N}}
\def\curlyNt{\tilde{\mathcal{N}}}
\def\gammad{\dot{\gamma}}
\newcommand{\fbar}{\bar{f}}
\newcommand{\kbar}{\bar{k}}
\newcommand{\sFrac}[2]{{\textstyle\frac{#1}{#2}}}
\newcommand{\OO}{\mathcal{O}}
\newcommand\Gbo{\overline{\Gb}}

\def\Gb{\bm{G}}

\def\kpa{k^\parallel}
\def\kpe{k^\perp}
\def\Gpa{G^\parallel}
\def\Gpe{G^\perp}
\def\p{\partial}

\renewcommand{\theequation}{S\arabic{equation}}
\renewcommand{\thefigure}{S\arabic{figure}}
\renewcommand{\bibnumfmt}[1]{[S#1]}
\renewcommand{\citenumfont}[1]{S#1}


This Supporting Information is organized as follows: in Section~\textbf{A} we provide details about the glass-forming models we employed in this work, and the preparation protocol used to generate our ensemble of glassy samples. In Section~\textbf{B} the first-order expansion in temperature of an interaction energy is derived, from which the definition of a LTE $\curlyE_\alpha$ emerges. We further explain how we calculate LTEs numerically, and discuss the generality of our results. In Section~\textbf{C} we present distributions of the magnitude of forces between particles in our model glass. In Section~\textbf{D} we describe how the LTE field is processed to give rise to soft spots and to predictions of ensuing plastic instabilities under shear. In Section~\textbf{E} we explain how we quantify the level of predictiveness of the LTE field and describe how soft spot maps based on a normal-mode analysis are constructed.

\section*{A.~Models and Preparation Protocols}

{\bf {\em Models} ---} We employ a single glass-forming model in two-dimensions (2D), and two glass-forming models in 3D, referred to as the 2DIPL, 3DIPL, and 3DKABLJ systems, respectively. The 2DIPL model is a 50:50 binary mixture of `large' and `small' particles of equal mass $m$, interacting via radially-symmetric purely repulsive inverse power-law pairwise potentials, that follow
\begin{equation}
\varphi(r_{ij}) = \left\{ \begin{array}{ccc}\epsilon\left[ \left( \sFrac{\lambda_{ij}}{r_{ij}} \right)^n + \sum\limits_{\ell=0}^q c_{2\ell}\left(\sFrac{r_{ij}}{\lambda_{ij}}\right)^{2\ell}\right]&,&\sFrac{r_{ij}}{\lambda_{ij}}\le x_c\\0&,&\sFrac{r_{ij}}{\lambda_{ij}}> x_c\end{array} \right.,
\label{eq:IPL}
\end{equation}
where $r_{ij}$ is the distance between the $i^{\mbox{\tiny th}}$ and $j^{\mbox{\tiny th}}$ particles, $\epsilon$ is an energy scale, and $x_c$ is the dimensionless distance for which $\varphi_{\mbox{\tiny IPL}}$ vanishes continuously up to $q$ derivatives. Distances are measured in terms of the interaction lengthscale $\lambda$ between two `small' particles, and the rest are chosen to be $\lambda_{ij}\!=\!1.18\lambda$ for one `small' and one `large' particle, and $\lambda_{ij}\!=\!1.4\lambda$ for two `large' particles. The coefficients $c_{2\ell}$ are given by
\begin{equation}
c_{2\ell} = \frac{(-1)^{\ell+1}}{(2q-2\ell)!!(2\ell)!!}\frac{(n+2q)!!}{(n-2)!!(n+2\ell)}x_c^{-(n+2\ell)}\,.
\end{equation}
We chose the parameters $x_c\!=\!1.48, n\!=\!10$, and $q\!=\!3$. The density was set to be $N/V\!=\!0.86\lambda^{-2}$; this choice sets the scale of characteristic $T\!=\!0$ interaction energies to be of order unity. We emphasize that this model glass-former does not lie in the proximity of an unjamming point since it possesses an intrinsic invariance to variations of density (or pressure), as established by the extensive work of Dyre et al.~\cite{jeppe_2008,jeppe,jeppe_nature_com}. Indeed, none of the observables measured in our simulations or in our analysis depend on our particular choice of density. The 2DIPL model undergoes a computer-glass-transition at a temperature of $T_g\!\approx\!0.5\epsilon/k_B$ for the density we chose.

The 3DIPL model is the three-dimensional version of the 2DIPL. Here we follow the same reasoning in setting the density and choose $N/V\!=\!0.82\lambda^{-3}$. The resulting glass transition temperature is $T_g\!\approx\!0.52\epsilon/k_B$.

The 3DKABLJ is the canonical Kob-Andersen binary Lennard-Jones model \cite{kablj}. It is a binary mixture of 80\% type A particles and 20\% type B particles of equal mass $m$, interacting via the following radially-symmetric pairwise potential
\begin{equation}
\varphi(r_{ij}) = \left\{\!\! \begin{array}{ccc}\varphi_{\mbox{\tiny LJ}}\!\!\left( \sFrac{r_{ij}}{\lambda_{ij}} \right) + \epsilon_{ij}\!\sum\limits_{\ell=0}^3 c_{2\ell}\left(\sFrac{r_{ij}}{\lambda_{ij}}\right)^{2\ell}&,&\sFrac{r_{ij}}{\lambda_{ij}}\le x_c\\0&,&\sFrac{r_{ij}}{\lambda_{ij}}> x_c\end{array} \right.,
\end{equation}
where $\varphi_{\mbox{\tiny LJ}}\!\!\left( \frac{r_{ij}}{\lambda_{ij}} \right)\!=4\epsilon_{ij}\!\left[\left( \frac{r_{ij}}{\lambda_{ij}} \right)^{12}\!\!-\left( \frac{\lambda_{ij}}{r_{ij}} \right)^6\right]$ is the conventional Lennard-Jones potential. Energies are expressed in terms of the A-A interaction $\epsilon\!\equiv\! \epsilon_{_{AA}}$, then $\epsilon_{_{AB}}\!=\!1.5\epsilon$ and $\epsilon_{_{BB}}\! =\! 0.5\epsilon$. The interaction length parameters are expressed in terms of $\lambda\! \equiv\! \lambda_{_{AA}}$, then $\lambda_{_{AB}}\! =\! 0.8$ and $\lambda_{_{BB}}\! =\! 0.88$. $x_c\! =\! 2.5$ is the dimensionless distance for which $\varphi$ vanishes continuously up to three derivatives. This condition sets the values of the coefficients $c_0\! = \!0.322042855424$, $c_2\!=\!-0.11564551766016$, $c_4\!=\!0.014774794872422$ and $c_6\!=\!-0.0006556954772111$. The density was set at $N/V=1.2$. With this parameter set the system undergoes a computer glass transition at $T_g\approx 0.45\epsilon/k_B$.

{\bf {\em Preparation protocol} ---} We prepared ensembles of glassy samples for all three models using the following protocol: first, systems were equilibrated in the high temperature liquid phase at $T\!=\!1.0\epsilon/k_B$. Then, the temperature was instantaneously set to a target value just below the respective $T_g$ of each model, where the dynamics were ran for a duration $t_{\mbox{\tiny anneal}}\!=\! 200\tau_0, 250\tau_0$ and $50\tau_0$ for the 2DIPL, 3DIPL and 3DKABLJ, respectively. Here $\tau_0\!\equiv\!\sqrt{m}\lambda/\epsilon$ is the microscopic units of time. This short annealing step is necessary to avoid generating unphysical ultra-unstable glassy configurations that could occur in an instantaneous quench, and is computationally advantageous compared to a continuous quench at a fixed quench-rate. After the annealing step we minimized the energy to produce glassy samples by a standard conjugate gradient method. Using this protocol, we have generated 5000 independent glassy samples for all three models, with $N\!=\!10000$ for the 2DIPL system, and $N\!=\!2000$ for the 3DIPL and 3DKABLJ systems.

\section*{B.~Local thermal energies}

In most of our work we omit particle indices with the goal of improving the clarity and readability of the text. We denote $N\dbar$-dimensional vectors as $\B v$, each component pertains to some particle index (e.g.~$i$) and some Cartesian spatial component (e.g.~$\xi$). Single, double and and triple contractions are denoted with $\cdot,:$, and $\tripleCdot$, respectively. For example, the notation $\frac{\partial^3{\cal A}}{\partial{\B x}\partial{\B x}\partial{\B x}}\!\tripleCdot\!{\B x}{\B x}{\B x}$ should be interpreted as $\sum_{ijk \xi \nu\upsilon}\frac{\partial^3{\cal A}}{\partial x_{i\xi}\partial  x_{j\nu} \partial x_{k\upsilon}}x_{i\xi}x_{j\nu}x_{k\upsilon}$, where $i,j,k$ run over particle indices and $\xi,\nu,\upsilon$ run over Cartesian spatial components.

We begin with deriving an expression for the thermal average of a general observable ${\cal A}\! =\! {\cal A}(\B x)$ which depends on the coordinates $\B x$, defined here as the displacement about an inherent state configuration. We denote with the superscript `(0)' quantities evaluated at the inherent state $\B x\!=\!0$ (i.e.~at zero temperature), e.g.~${\cal A}^{(0)}$, and $\C U(\B x)$ denotes the potential energy.

The mean of the observable ${\cal A}$ is a function of temperature, defined as
\begin{equation}\label{foo00}
\langle {\cal A} \rangle_{_T} \!\equiv\! \frac{\int\!\!{\cal A}(\B x)\exp\!\left(\!-\frac{{\C U}(\B x)}{k_BT}\!\right)d{\B x}}{\int\!\!\exp\!\left(\!-\frac{{\C U}(\B x)}{k_BT}\!\right)d{\B x}} \!=\! \frac{\int\!\!{\cal A}(\B x)\exp\!\left(\!-\frac{\delta{\C U}(\B x)}{k_BT}\!\right)d{\B x}}{\tilde{\cal Z}(T)} \,,
\end{equation}
where $\delta\C U\!\equiv {\C U} - {\C U}^{(0)}$ is the energy variation about the inherent state energy ${\C U}^{(0)}$, and $\tilde{\cal Z}(T)\!\equiv\! \int \exp\!\left(\!-\frac{\delta{\C U}(\B x)}{k_BT}\!\right)d{\B x}$ is the relevant partition function. $\delta\C U$ is expanded to third order in the coordinates as
\begin{equation}
\delta\C U \simeq \sFrac{1}{2}{\B {\C M}}\!:\!{\B x}{\B x} + \sFrac{1}{6}{\B {\C U}}'''\!\tripleCdot\!{\B x}{\B x}{\B x}\,, \\
\end{equation}
where ${\B {\C M}}\!\equiv\!\frac{\partial^2\C U}{\partial{\B x}\partial {\B x}}$ is the dynamical matrix, and ${\B {\C U}}'''\!\equiv\!\frac{\partial^3\C U}{\partial{\B x}\partial{\B x}\partial{\B x}}$ is the third-order tensor of derivatives of the potential energy. In what follows we assume that the scale of characteristic fluctuations of the coordinates is set by the equipartition theorem, namely $\langle x^2 \rangle\! \sim\! k_BT$, and therefore higher order products of coordinates are much smaller than $k_BT$. With this assumption, we expand the numerator of Eq.~(\ref{foo00}) as
\begin{widetext}
\begin{eqnarray}
\int {\cal A}(\B x)\exp\!\left(\!-\frac{\delta{\C U}(\B x)}{k_BT}\!\right)d{\B x} & \simeq & \int \left({\cal A}_0 + \frac{\partial {\cal A}}{\partial\B x}\!\cdot\!{\B x} + \frac{1}{2}\frac{\partial^2{\cal A}}{\partial{\B x}\partial{\B x}}\!:\!{\B x}{\B x}\right)\exp\left(-\frac{{\B {\C M}}\!:\!{\B x}{\B x}}{2k_BT}\right)\left(1 - \frac{{\B {\C U}}'''\!\tripleCdot\!{\B x}{\B x}{\B x}}{6k_BT}\right)d\B x\, \nonumber \\
& \simeq & \int \left({\cal A}_0 + \frac{1}{2}\frac{\partial^2{\cal A}}{\partial{\B x}\partial{\B x}}\!:\!{\B x}{\B x} - \frac{1}{6k_B T}\B x\!\cdot\!\frac{\partial {\cal A}}{\partial\B x}\, {\B {\C U}}'''\!\tripleCdot\!{\B x}{\B x}{\B x}\right)\exp\left(-\frac{{\B {\C M}}\!:\!{\B x}{\B x}}{2k_BT}\right)d\B x   \\
& = & \left( {\cal A}_0 + \frac{k_BT}{2}\left[\frac{\partial^2{\cal A}}{\partial{\B x}\partial{\B x}}\!:\!{\B {\C M}}^{-1} - \frac{\partial {\cal A}}{\partial\B x}\!\cdot\!{\B {\C M}}^{-1}\!\cdot{\B {\C U}}'''\!:\!{\B {\C M}}^{-1}\right]\right)\int\exp\left(-\frac{{\B {\C M}}\!:\!{\B x}{\B x}}{2k_BT}\right)d\B x\ ,\nonumber
\end{eqnarray}
where we have used the identities
\[
\int {\B x_i} {\B x_j} \exp\left(-\frac{{\B {\C M}}\!:\!{\B x}{\B x}}{2k_BT}\right)d\B x = T {\B {\C M}}_{ij}^{-1}\int \exp\left(-\frac{{\B {\C M}}\!:\!{\B x}{\B x}}{2k_BT}\right)d\B x\,,
\]
\[
\int {\B x_i} {\B x_j} {\B x_k} {\B x_m} \exp\left(-\frac{{\B {\C M}}\!:\!{\B x}{\B x}}{2k_BT}\right)d\B x = T^2\left( {\B {\C M}}_{ij}^{-1}{\B {\C M}}_{km}^{-1} + {\B {\C M}}_{ik}^{-1}{\B {\C M}}_{jm}^{-1} + {\B {\C M}}_{im}^{-1}{\B {\C M}}_{jk}^{-1}\right)\int \exp\left(-\frac{{\B {\C M}}\!:\!{\B x}{\B x}}{2k_BT}\right)d\B x\,.
\]
\end{widetext}
Since $\tilde{\cal Z}(T) \simeq \big(1+{\cal O}(k_BT)\big)\int \!\exp\!\left(\!-\frac{{\B {\C M}}:{\B x}{\B x}}{2k_BT}\!\right)d\B x$, we arrive at the result
\begin{equation}\label{foo01}
\frac{\langle {\cal A} \rangle_{_T} - {\cal A}^{(0)}}{\frac{1}{2}k_BT} \simeq \frac{\partial^2{\cal A}}{\partial{\B x}\partial{\B x}}\!:\!{\B {\C M}}^{-1} - \frac{\partial {\cal A}}{\partial\B x}\!\cdot\!{\B {\C M}}^{-1}\!\cdot{\B {\C U}}'''\!:\!{\B {\C M}}^{-1}\,.
\end{equation}
as appears in the main text. We stress that the effect of higher order derivatives of both ${\cal A}$ and ${\C U}$ can also be explicitly calculated and is of a higher order in $T$ (not shown).

In this work we study the local thermal energy (LTE), defined as follows: we focus on potential energy functions that can be written as a sum over pairwise interactions ${\C U}\! =\! \sum_\alpha \varepsilon_\alpha$, where $\alpha$ labels the different pairs of interacting degrees of freedom. Using Eq.~(\ref{foo01}), we define the local thermal energy $\curlyE_\alpha$ as
\begin{eqnarray}
{\C E}_\alpha & \equiv & \lim_{T \to 0}\!\frac{\langle{\varepsilon}_\alpha\rangle_{_T}-{\varepsilon}_\alpha^{(0)}}{\tfrac{1}{2}k_B T} \nonumber \\
& = & \frac{\partial^2\varepsilon_\alpha}{\partial \B x\partial \B x}\!:\!{\B {\C M}}^{-1} - \frac{\partial\varepsilon_\alpha}{\partial {\B x}}\!\cdot\!{\B {\C M}}^{-1}\!\cdot{\B {\C U}}'''\!:\!{\B {\C M}}^{-1}\,. \label{foo02}
\end{eqnarray}

Examples of the LTE fields calculated in 2D model glasses can be found in Figs.~1 and 3 in the main text. These fields are calculated as follows: we perform a full diagonalization of the dynamical matrix ${\B {\C M}}$ calculated for each glassy sample, and obtain the complete set of eigenmodes $\{{\B \Psi}_\ell\}_{\ell=1}^{N\dbar}$ and their associated eigenfrequencies $\{\omega_\ell\}_{\ell=1}^{N\dbar}$, where $\dbar$ is the spatial dimension. We then solve the following linear equation for the thermal displacements ${\B {\C X}}\!\equiv\!-{\B {\C M}}^{-1}\!\cdot{\B {\C U}}'''\!:\!{\B {\C M}}^{-1}$ (see main text) using a conventional conjugate gradient solver
\begin{equation}\label{foo04}
{\B {\C M}}\!\cdot\!{\B {\C X}} = -\sum_{\ell}\frac{{\B {\C U}}'''\!:\!{\B \Psi}_{\ell}{\B \Psi}_{\ell}}{\omega_\ell^2}
\end{equation}
Expressions for ${\B {\C M}}$ and ${\B {\C U}}'''$ for pairwise potentials are available in e.g.~\cite{athermal_elasticity}. Finally, the LTE $\curlyE_\alpha$ is calculated for each interaction $\alpha$ as
\begin{equation}\label{foo05}
\curlyE_\alpha = \sum_\ell \frac{\frac{\partial^2 \varepsilon_\alpha}{\partial \B x\partial \B x}\!:\!{\B \Psi}_{\ell}{\B \Psi}_{\ell}}{\omega_\ell^2} + \frac{\partial \varepsilon_\alpha}{\partial \B x}\!\cdot\!{\B {\C X}}\,.
\end{equation}

\begin{figure}[ht]
\centering
\includegraphics[width=.9\linewidth]{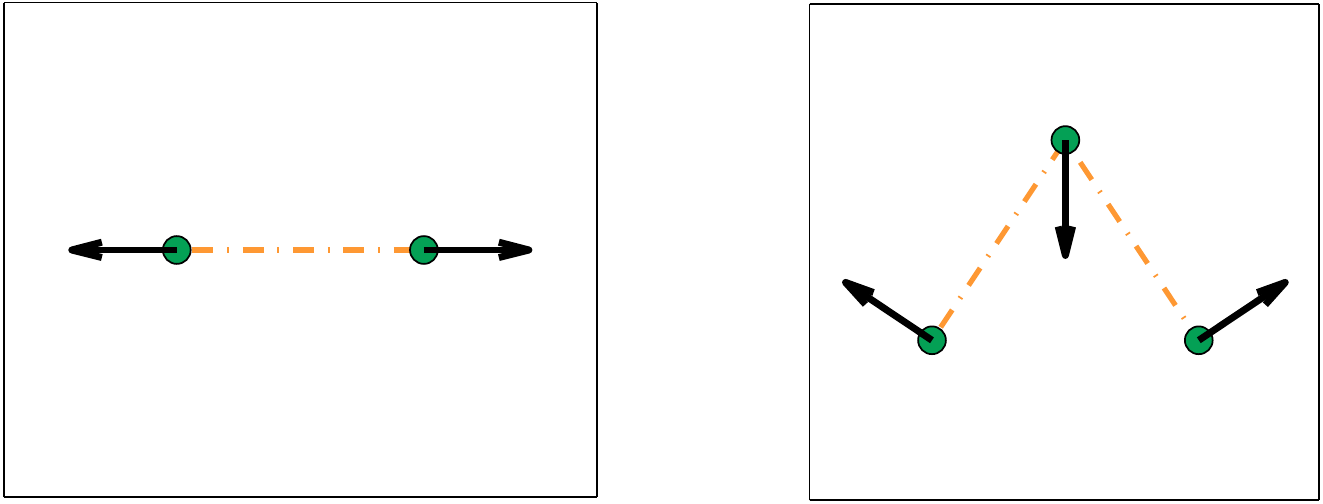}
\caption{Left panel: the geometry of ${\B f}_\alpha$ in the case that $\varepsilon_\alpha$ is a pairwise radially-symmetric interaction. Right panel: same as left panel, for the case of a 3-body $\varepsilon_\alpha$ which depends on the angle between the triple of nodes.}
\label{fig:force_geometry}
\end{figure}

The formalism presented above remains valid for systems in which the potential is written as a sum of 3-body (or higher) terms, e.g.~\cite{Stillinger_Weber}. In this case ${\C U}\!=\!\sum_\alpha \varepsilon_\alpha$, where now $\alpha$ labels a \emph{triple} of interacting particles. The same expression given in Eq.~(\ref{foo02}) would now describe the LTE associated with the triple $\alpha$. A key point is that the forces
\begin{equation}
{\B f}_\alpha \equiv \frac{\partial \varepsilon_\alpha}{\partial \B x}
\end{equation}
have a different form in the case of 3-body interactions compared to the case of pairwise interactions. In the latter, if the interaction is radially-symmetric, ${\B f}_\alpha$ has the geometry of a dipole vector acting on the pair $\alpha$, as illustrated in the left panel of Fig.~\ref{fig:force_geometry}. What is the form of ${\B f}_\alpha$ for 3-body interactions? As an example, assume that the interaction $\varepsilon_\alpha\!=\!\varepsilon_\alpha(\theta_\alpha)$ depends upon the angle $\theta_\alpha$ formed between a triple $i,j,k$ of particles. In this case, ${\B f}_\alpha$ is a field with the geometry as illustrated in the right panel of Fig.~\ref{fig:force_geometry}.

We assert that as long as the interaction potential is translationally and rotationally invariant (i.e.~it only depends on the relative distances and orientations between the triple $\alpha$), the associated ${\B f}_\alpha$ will be of a form which, when contracted with a slowly-varying field in space, will pick up contributions that are proportional to the spatial gradient of the slowly-varying field. The same reasoning also applies to contractions of slowly-varying fields with the third-order tensor ${\B {\C U}}'''$ as well. For these reasons, we expect LTEs to always filter out collective translational modes, and therefore be insensitive to the presence of low-frequency plane-waves, independently of the particular form of the potential energy.

Finally, we comment on the computational complexity of our numerical analysis: the bottleneck of the calculation is the requirement to obtain all the eigenmodes and eigenvalues of the dynamical matrix. The computational time of this full-diagonalization is known to scale as $N^3$. The computational time dedicated to the rest of the analysis is negligible compared to the diagonalization step. It is left for future research to investigate whether a partial diagonalization of the dynamical matrix (which would simply result in truncated sums in Eqs.~(\ref{foo04}) and (\ref{foo05})) would suffice for producing softness maps with comparable predictive powers to those obtained using a full diagonalization.

\section*{C.~Distribution of force magnitudes}

\begin{figure}[ht]
\centering
\includegraphics[width=.85\linewidth]{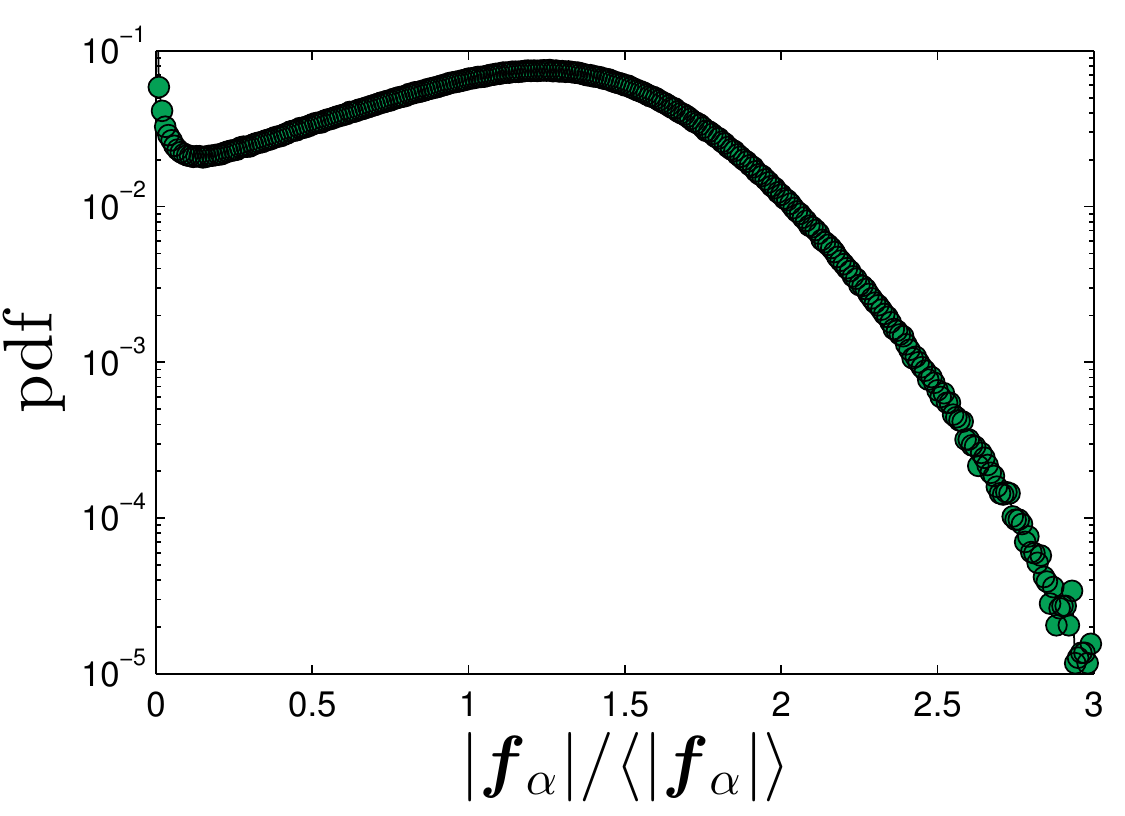}
\caption{The distribution of the magnitude of forces between particles measured for the 3DIPL system shows a superexponential decay at large values.}
\label{fig:force_distribution}
\end{figure}

In the main text we present a scaling argument according to which the distribution of LTEs should follow $p(\curlyE_\alpha)\!\sim\!\curlyE_\alpha^{-9/4}$, based on the recent discovery that the asymptotic form of the distribution of glassy low-frequency modes in glassy systems follows $D_G(\omega)\!\sim\!\omega^4$~\cite{modes_prl}. In this argument, we assume that the magnitudes of forces between the glass particles is narrowly distributed. Here, we present numerical evidence that validates this assumption: in Fig.~\ref{fig:force_distribution} we present the distribution of the magnitude of pairwise forces between particles in the 3DIPL system, showing that it decays superexponentially at large values.

\section*{D.~Identifying soft spots and their degree of softness}

To quantitatively analyze the heterogenous spatial distribution of LTE ${\C E}_\alpha$, we construct coarse-grained 2D map as follows. Space is discretized into bins of the smallest size which proved to always include at least two bonds' center of masses. In our case, it corresponds to a bin size of $1/40 L$, with $L$ being the linear size of the simulation box. The coarse-grained map is then built in two steps. In the first step, each bond is associated with a bin selected according to the bond's center of mass and the absolute value of its LTE contributes to the bin's value. In the second step, the map is smoothed out by averaging the bin's value with the values of all bins in the first layer of neighboring bins ($8$ bins in 2D). For easier processing, bonds with an associated LTE value smaller than $|{\C E}_\alpha|\!=\!1.1$ were omitted. We verified that this choice does not affect the results, which are sensitive to large values of $|{\C E}_\alpha|$.

The local maxima of the coarse-grained map were then extracted. These maxima are identified with soft spots, as described next. We first analyzed each row of the 2D maps at a time, where the bins corresponding to a local maximum were flagged. We repeated the same flagging procedure for every column. Bins which were flagged twice were defined as soft spots. As the exact location of the soft spot within the bin's area is of no interest, we define the soft spot location as the bin's coordinate with added white noise to avoid discretization effects. We used the bin's value as the soft spot score $\hat\eta$, which describes the average value of $|\curlyE_\alpha|$ in the near vicinity of the soft spot center.

The LTE of bonds are widely distributed and consequently so are the scores of the soft spots, both within and between realizations. We therefore adopt the following standardized score
\begin{equation}
\Delta_\curlyE=\frac{\eta_{\rm max}}{\hat\eta_i} \ ,
\end{equation}
where $\hat\eta_i$ is the score of the $i^{\rm th}$ soft spot and $\eta_{\rm max}\!=\!{\rm max}_i\left[{\hat\eta_i}\right]$ evaluated for each realization. Therefore, the softest spot in each realization has $\Delta_\curlyE\!=\!1$ and not-as-soft spots are characterized by $\Delta_\curlyE\!>\!1$, where the deviation from unity quantifies the degree of softness within each realization. This standardization allows a consistent numerical analysis per realization, as well as the calculation of distribution functions based on a large number of realizations. The analysis is based on $5000$ independent realizations, where a few tens of soft spots were detected per realization. Among these spots, the softest ones --- i.e.~those with $\Delta_\curlyE$ close to unity --- dominate the plastic response under shearing. For example, there are on average $25$ spots with $\Delta_\curlyE\!\le\!2$, which according to Fig.~4 (left), predict nearly $70\%$ of the first plastic events.

\section*{E.~quantifying predictiveness of plastic rearrangements under shear}

\begin{figure}[ht]
\centering
\includegraphics[width=\linewidth]{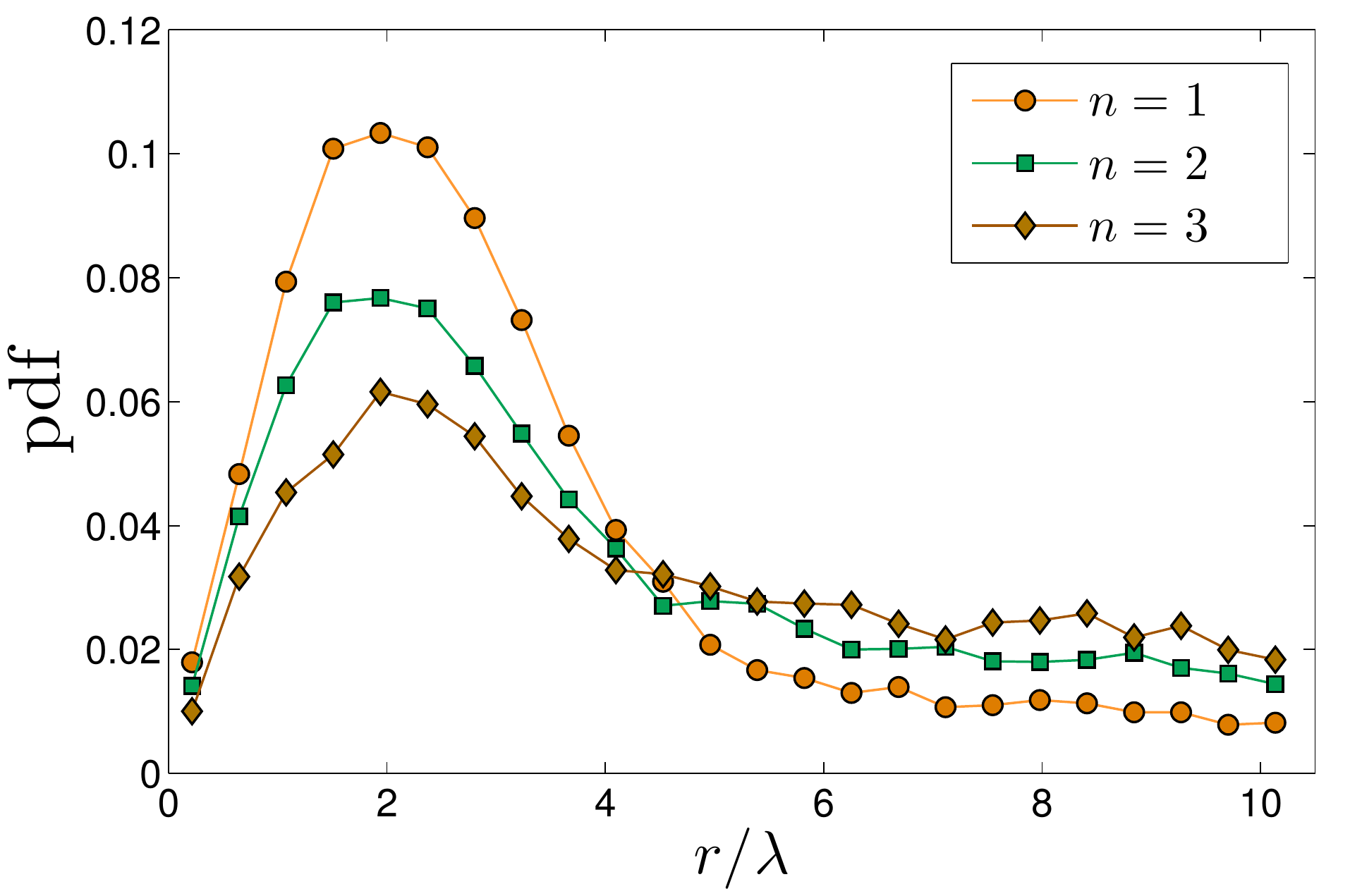}
\caption{The probability distribution function of the distance $r$ (normalized by the bond-length between two small particles $\lambda$, cf.~Eq.~\eqref{eq:IPL} and the text below it) between plastic events and the center of the nearest soft spot. It is observed that plastic events occur with high-probability close --- within a few bond-lengths --- to soft spots.}
\label{fig:eventCloseness}
\end{figure}
{\bf {\em Plastic rearrangements} ---} The performed athermal quasi-static shearing simulations followed well-established two-step protocols of first imposing an affine simple-shear transformation to the system and then minimizing its energy while enforcing Lees-Edwards boundary conditions, see e.g.~\cite{lemaitre2004_avalanches,lemaitre2006_avalanches,steady_states_with_jacques}. During these simulations the energy was used as an indicator of plastic rearrangements/events which were identified with strain precision up to $10^{-6}$ using backtracking methods. The plastic events were automatically spatially localized by selecting the particle with the largest displacement value as a consequence of the energy minimization step at the occurrence of the plastic event.

{\bf {\em Quantification} ---} Each glass realization was sheared until $5$ plastic events were triggered. The probability distribution function of the distance of plastic events to the nearest spot in space is shown in Fig.~\ref{fig:eventCloseness}. It is observed that plastic events occur with high-probability near soft spots (corresponding to the peak around $2$ bond-lengths). Consequently, we identify the soft spot which is closest to the $k^{\rm th}$ plastic event and record its standardized score $\Delta_{\curlyE}$ for further analysis as described in the manuscript.

{\bf {\em Normal-modes maps} ---} To compare the LTE results with existing methods/results in the literature, we followed the protocol described in~\cite{widmer2008irreversible} to produce a field which is based on the $30$ lowest normal-modes with non-vanishing associated energy. By constructing such normal-modes-based maps, each and every particle in the system has a score corresponding to a sum over the displacement squared of the modes. We then applied exactly the same protocol described in Section~\textbf{D} in the context of the LTE maps to the normal-modes-based maps, where bonds' centers of mass were replaced with particle positions and the LTE absolute values of bonds were replaced with particles' scores. The results of the comparison are presented in the main text.


\end{document}